# Application of laser biospeckle analysis for assessment of seed priming techniques

**Puneet Singh[1], Amit Chatterjee[1], Vimal Bhatia[1] and Shashi Prakash[2*]**

[1]Signals and Software Group, Discipline of Electrical Engineering, Indian Institute of Technology Indore-453552, India.

[2]Photonics Laboratory, Department of Electronics & Instrumentation Engineering, Institute of Engineering & Technology, Devi Ahilya University, Khandwa Road, Indore-452001, India.

*Corresponding Author,
Phone: +91 731 2361116, 09977186156,
Fax: 91 731 2764385,
E-mail: sprakash_davv@rediffmail.com

**ABSTRACT:**

Seed priming is one of the well-established and low cost method to improve seed germination properties, productivity, and stress tolerance in different crops. It is a pre-germination treatment that partially hydrates the seed and allows controlled imbibition. This stimulates and induces initial germination process, but prevents radicle emergence. Consequently, treated seeds are fortified with enhanced germination characteristics, improved physiological parameters, uniformity in growth, and improved capability to cope up with different biotic and abiotic stresses. Existing techniques for evaluating the effectiveness of seed priming suffer from several drawbacks, including very high operating time, indirect and destructive analysis, bulky experimental arrangement, high cost, and require extensive analytical expertise. To circumvent these drawbacks, we propose a biospeckle based technique to analyse the effects of different priming treatments on germination characteristics of seeds. The study employs non-primed ($T_0$) and priming treatments ($T_1$-$T_{75}$), including hydropriming (6 hours, 12 hours, 18 hours, and 24 hours) and chemical priming (using three chemical agents namely sodium chloride, potassium nitrate, and urea) for different time durations (6 hours, 12 hours, 18 hours, and 24 hours) and solution concentrations (1%, 2%, and 5%). To analyse the effect of priming using biospeckle analysis, the treated seed samples were irradiated by a spatially filtered laser beam. The resultant dynamic speckle patterns are successively captured using a CCD camera. The non-normalized histogram based robust indexing method in conjunction with Gaussian curve fitting has been used on the captured images to assess both, qualitative and quantitative mapping of dynamicity. An increase in the biospeckle activity of the seed is observed after priming. This is due to the enhance level of certain physiological and biochemical processes related to germination process, occurring inside the seed. Obtained results are then correlated with the data acquired from the standard germination test. The results conclusively establish biospeckle analysis as an efficient active tool for seed priming analysis. Furthermore, the proposed setup is extremely simple, low-cost, involves non-mechanical scanning and is highly stable.

# 1. INTRODUCTION

Seed quality is one of the most important factor, necessary to meet the current demand for high quality yield. It ensures high germination rate and production uniformity. Germination is the most important phase in the stand establishment of a crop and is directly related to the quality of the seed [1]. Seed germination is the fundamental stage in a plant's lifecycle and depends on the environmental conditions. Seed should germinate rapidly and emerge homogenously so that the light, water, and soil minerals can be used effectively. Speedy and consistent transition towards the field emergence is an essential prerequisite to attain the optimum yield potential, quality, and eventually profit in different annual crops. Therefore, different techniques have been introduced to improve stand establishment of the plants for enhancing the production rate under normal as well as in a stressful environment [2]. Plants are exposed to potentially adverse biotic and abiotic stresses, which subsequently alters the growth and productivity of the plants. Nowadays, different strategies are being introduced to produce stress-tolerant plants. Environmental stresses are induced due to various reasons including the climate changes, resulting in exposing crops to intermittent or terminal drought conditions, high salinity, extreme temperature, and submergence, which leads to impairment of plant development. These distressful situations may also result in the premature death of the plant.

As the seed quality is the fundamental and crucial parameter for improving productivity and profitability, hence extensive investigations have been undertaken to improve seed quality. In recent years, priming has evolved as an efficient tool for enhancing seed quality, and produce stress tolerant plants. Priming is one of the well-established and extremely low cost method for seed quality enhancement [3]. It helps in enhancing the germination capability of seeds and stand establishment by altering seed vigor and/or the physiological state of the seed. The fast transition of seed towards germinated state is the result of enhancement in germination metabolism, restoring process, and improved antioxidant activity [4]. The priming improves stand establishment, germination rate, seedling vigour, vigour index, viability, and uniformity of growth [5].

In priming, seeds are exposed to external water potential under the controlled conditions which allows some of the pre-germinative physiological and biochemical processes to occur before germination takes place. By regulating germination processes before sowing the seed, priming increases the germination rate and at the same time, it also accelerates the germination speed. Various priming methods have been used for increasing the germination properties of seeds. Seeds were immersed in sterilized distilled water at appropriate temperature for specific duration (hydropriming); different salts and aerated low-water-potential solutions (osmopriming); mixing of seeds with solid or semi-solid material and providing specific amount of water slowly (solid matrix priming); coating seeds with formulated product of bio-agents after hydrating them with water (bio-priming), etc. [6]. Several studies have been conducted to analyse the effect of priming on germination rate, field emergence, seedling vigor, stand establishment, stress resistivity, and economic yields in different crops [7]. In this regard, Anisa et al. [8] reported the effect of seed priming on the pattern of seed imbibition and germination with different concentration of potassium nitrate. Priming is also regularly used for the different vegetable seeds including carrot, onion and leek, celery, tomato, and others, for enhancing overall seed germination properties. It has also been very effective for improving the product quality of seeds in the flower industry [9]. Waller et al. [10] reported the effect of seed priming on seedling growth and germination under drought stress. The study in [10] concluded that the seed priming can increase the drought tolerance of Chinese cabbage during germination.

Prevailing agricultural science and technology can provide basic pre-harvest monitoring, and can analyse the fundamental causes of decreasing productivity in different crops. However, further advancement in the methods for the analysis of seed quality enhancement and different pre-treatments based on multidisciplinary approaches are necessary to ensure the optimum yield. In the past, few strategies like incubator/germinator based germination, sowing the treated seeds were used for analysing the effect of priming on different seeds. However, these methods for analysis are slow, destructive, expensive, manual, and require extensive analytical and theoretical expertise. So far, no direct method is available to analyse the effect of different priming treatments without prior germination. Thus, a simple, fast, non-destructive and low cost device for seed priming analysis shall be very useful.

Recently, optical methods have emerged as important tools of analysis in agricultural sciences. These methods provide us a substitute for existing manual testing methods, as they are very fast and non-destructive. In this context, the dynamic laser speckle is one of the most promising technique. It is becoming very popular diagnostic tool due to its experimental simplicity, fast acquisition process and robust analysis. Biospeckle is a phenomenon that occurs when an object having physical and chemical activity at microscopic level is irradiated by a coherent source [11]. So far, laser biospeckle has been used for wide range of applications in several fields, like agriculture, food engineering, pomology, microbiology, medicine, biomedical engineering, cancer detection. Specifically, researchers have investigated biometric spoofing [12, 13], bruising in fruits (maturity, damage, aging, mechanical properties) with its biospeckle behaviour [14], fungal and bacterial colony growth [15], biological activity in artificially cultured root tissues [16], kafir grain activity [17], beef spoilage with aging [18], bread spoilage by fungal contamination [19], breast cancer [20], seed viability [21], differentiation of homogeneous and heterogeneous seeds [22], quality assessment of apple fruit [23], paint drying dynamics [24], pre- and post-harvest phenomenon (e.g. chlorophyll content, mealiness, temperature effect, disease detection) related to apple fruit [25, 26], etc. using the same principle.

In this work, we develop a biospeckle based sensor for characterization of seed priming treatments. The setup used for biospeckle analysis is very simple, low cost and is capable of analysing the samples with high accuracy, within a short time duration. The non-normalized histogram based method combined with the Gaussian fitting function has been used to get visual and numerical assessment of the activity for different samples.

Rest of the paper is organized as follows: In Section 2, the biospeckle phenomenon and the mathematical description of the analysis is given. Section 3 describes sample preparation and experimental procedure for the priming analysis. We have summarized the results in Section 4. Lastly, conclusions and inferences are drawn from the biospeckle analysis in Section 5.

## 2. Theory

Speckle pattern is generated when coherent light is either reflected from a rough surface or propagates through a medium having temporal variations [11]. These variations are generally observed in biological objects and results in time varying statistical interference pattern. The light field at a particular point Q(x,y,z) in a speckle pattern is the sum of large number of components having contribution from all points in the scattering field. Mathematically, the speckle signal at point Q generated by any surface element k, is given by [27]:

$$u_k(Q) = |u_k|e^{j\emptyset_k} = |u_k|e^{jlr_k} \tag{1}$$

where, $j = \sqrt{(-1)}$

$l$= wave number.

$r_k$= random varying distance from k$^{th}$ scattering element to point Q.

When biological specimens (like fruits, blood, seeds), are irradiated by laser light, the resultant speckle pattern changes with respect to time, and is termed as biospeckle [27]. Over the last few decades, this technique has been extensively used for characterization of different properties (like viability, bruise, freshness, liveliness) of biological specimen. In order to assess the physiological or chemical activity of the specimen, M time series images I=[i$_1$,i$_2$,i$_3$,...,i$_M$] of size M$_x$×M$_y$ are captured at a sampling rate fs=T/M (where, T is the time duration for capturing all the frames). Modified structural function (MSF) has been used for obtaining the visual activity map, and histogram based non-normalized technique quantifies the statistical data [12]. The algorithm includes following steps:

I. MSF visualization: To generate 2D visual activity map, MSF method is used. The analysis is based on summation of absolute difference between pixels of images i$_j$ and i$_{j+m}$ separated by 'm' frames. This can be mathematically expressed as [12],

$$\text{MSF}(x,y) = \sum_{j=1}^{M-m}|i_j(x,y) - i_{j+m}(x,y)| \tag{2}$$

where, x and y are co-ordinates of intensity matrix i$_j$, m is the window length, j is sequence index and MSF(x,y) is the resulting 2D visual activity map.

II. Histogram mapping: To generate the probability distribution function (PDF) from a point wise distributed MSF(x,y), calculation of its non-normalized histogram map is performed. For a digital image, the histogram function (with gray levels in the range [0, L-1]) can be defined as:

$$h(r_k) = n_k \tag{3}$$

where, $r_k$ = $k^{th}$ gray level,

$n_k$ = number of pixels in the image having gray level $r_k$.

The MSF activity map should contain enough number of pixels to provide a consistent distribution of frequency counts of the estimate values that can be precisely represented through its PDF.

III. Gaussian curve fitting: As speckle is a random, naturally occurring phenomena, to fit the histogram map, Gaussian fitting using Caruana's algorithm [28] was employed. The process involves taking natural logarithm of the Gaussian histogram function:

$$\ln(y) = \ln(A) + \frac{-(x-\mu)^2}{2\sigma^2}$$

$$= \ln(A) - \frac{\mu^2}{2\sigma^2} + \frac{2x\mu}{2\sigma^2} - \frac{x^2}{2\sigma^2}$$

$$= a + bx + cx^2 \tag{4}$$

where, $a = \ln(A) - \frac{\mu^2}{2\sigma^2}$

$$b = \frac{\mu^2}{2\sigma^2}$$

and, $c = \frac{1}{2\sigma^2}$

The values of the indices (a, b and c) are calculated by differentiating the basis equation and setting the resultant expression to zero. This yields a linear system of equations:

$$\begin{bmatrix} N & \sum x & \sum x^2 \\ \sum x & \sum x^2 & \sum x^3 \\ \sum x^2 & \sum x^3 & \sum x^4 \end{bmatrix} \begin{bmatrix} a \\ b \\ c \end{bmatrix} = \begin{bmatrix} \sum \ln(y) \\ \sum x \ln(y) \\ \sum x^2 \ln(y) \end{bmatrix}$$

where, N = the number of data points in the observation and Σ is the sum over [1, N].

After solving Eq. (4), the values of mean ($\mu$) and standard deviation ($\sigma$) are obtained; which are the parameters of interest to fit the Gaussian curve to the scattered histogram points.

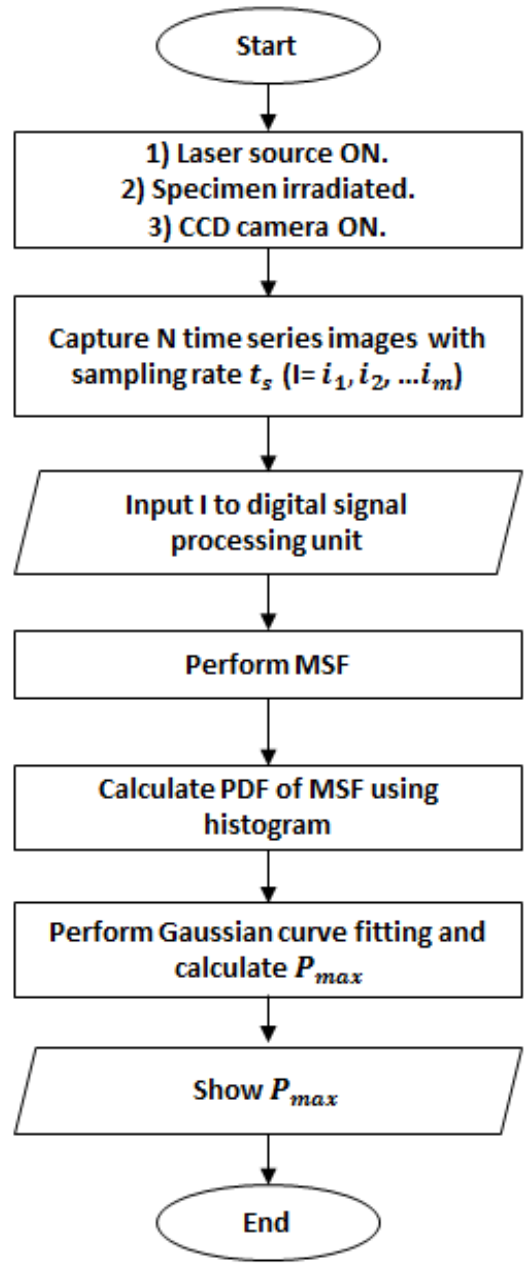

**Fig. 1. Flow diagram for the biospeckle processing algorithm.**

IV. Quantitative analysis: Quantitative information of statistically fitted data is extracted from Gaussian curve by tracing its non-normalized pixel level ($P_{max}$) corresponding to the maximum frequency counts. This step provides numerical characterization of dynamic speckle activity from the 2D visual result.

The overall flow chart representation of the algorithm used for biospeckle processing is shown in Fig. 1.

Researchers have anticipated biospeckle phenomenon as a result of various physiological and biochemical processes, like cytoplasmic streaming, cell divisions, biochemical reactions, organelle movement, Brownian motion, etc. However, the knowledge of biospeckle activity related to fruits, vegetables, leaves, crops is still limited and there is a lack of consistent biological interpretations of the phenomenon. In this paper, we have investigated the applicability of laser biospeckle for seed priming assessment. The correlation between biospeckle dynamicity with the extent of the seed germination characteristics for non-treated and treated seed has been successfully demonstrated.

### 3. Biospeckle measurement and sample preparation

3.1. Seed priming treatments

To exemplify the biospeckle method for evaluting the effect of priming chickpea *(Cicer arietinum L.)* seed has been used. It provides highly nutritious food and ranks third among the pulse crop in terms of the total production, which accounts to about 10.1 million tons per annum [29]. The poor germination characteristics of the seed under normal as well as in stressful environment can adversely affects the crop production. For experimentation native chickpea seed lots were procured locally. 8000 seeds were selected from the lot and divided into 40 slots (200 seeds /replicate).The chemical solutions (sodium chloride, potassium nitrate and urea) with different concentration (1%, 2% and 5% w/v) were prepared. Before the treatments, chickpea seeds were surface sterilized with distilled water. After that, washed seeds were fully submerged into the solutions for priming. These samples were placed in aseptic condition for different time periods (6, 12, 18 and 24 hours) at $25\pm1\ ^0C$ (temperature monitoring was performed using FLIR C2 thermal camera). Different priming treatments were applied, which are divided into two broad categories namely hydro and chemical priming treatments ($T_1$-$T_{75}$). Table 1 shows the summary of all the priming treatments used for the experimentations.

**Table 1**

Priming treatment applied to chickpea seeds.

| Priming Treatment | Treatment notation | Duration in hours (h) | Treatment agent | Concentration |
|---|---|---|---|---|
| Non-primed (Controlled) | $T_0$ | - | - | - |
| Hydropriming | $T_1$ | 6 | Distilled Water | - |
|  | $T_2$ | 12 | Distilled Water | - |
|  | $T_3$ | 18 | Distilled Water | - |
|  | $T_4$ | 24 | Distilled Water | - |
| Chemical priming | $T_{51}$ | 6,12,18,24 | Sodium Chloride | 1% |
|  | $T_{52}$ | 6,12,18,24 | Sodium Chloride | 2% |
|  | $T_{55}$ | 6,12,18,24 | Sodium Chloride | 5% |
|  | $T_{61}$ | 6,12,18,24 | Potassium Nitrate | 1% |
|  | $T_{62}$ | 6,12,18,24 | Potassium Nitrate | 2% |
|  | $T_{65}$ | 6,12,18,24 | Potassium Nitrate | 5% |
|  | $T_{71}$ | 6,12,18,24 | Urea | 1% |
|  | $T_{72}$ | 6,12,18,24 | Urea | 2% |
|  | $T_{75}$ | 6,12,18,24 | Urea | 5% |

3.2. Experimental setup

As the seed priming is completed, the seeds were cleaned with the sterilized distilled water and placed between filter paper for 10 min to remove excess water on the seed surface. The treated seeds were then placed in front of the biospeckle apparatus as shown in Fig. 2 (a) for dynamicity assessment.

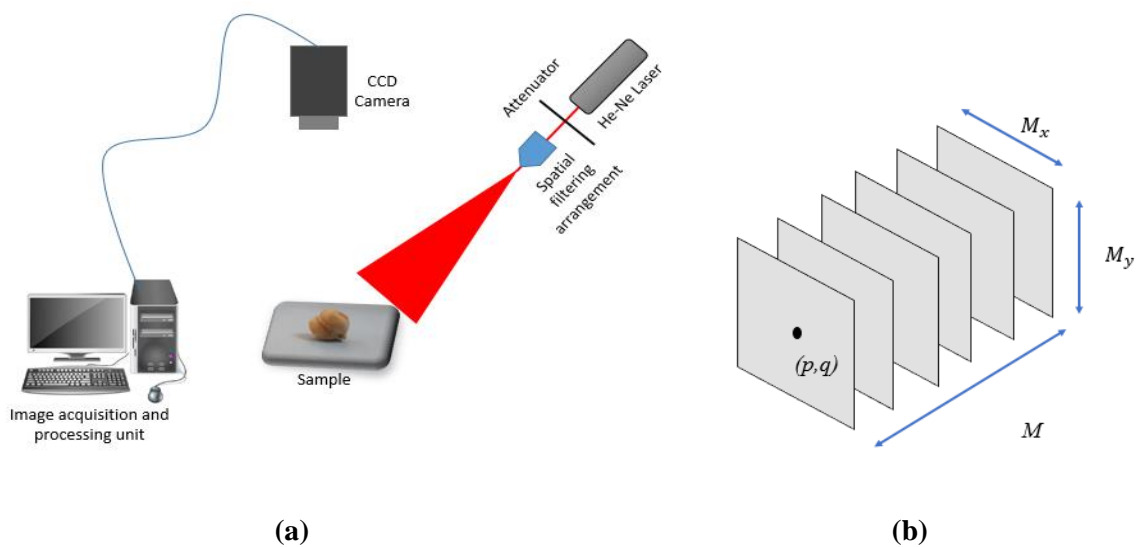

(a)  (b)

**Fig. 2 (a) Experimental setup for biospeckle recording, and (b) time sequence speckle images.**

The primed and nonprimed seeds were illuminated with He-Ne laser beam (15 mW, λ= 632.8 nm). The intensity of the incoming laser beam was controlled by the means of variable attenuator. The experiment was repeated multiple times and quality test protocol (QTP) [30] was followed to remove the possibility of false activity generation.

Spatial filtering arrangement consisting of microscopic objective (MO) of magnification 40X and an aperture of 10 microns was used to filter and expand the beam. The successive biospeckle images were recorded by a CCD camera (Basler Corp., frame rate: 30 fps, resolution: 1024×967) and converted into digital format using frame grabber card. The time frame sequence of these encoded images were created as shown in Fig. 2 (b), and analysed using MATLAB based analysis algorithm.

**4. Results and discussion**

4.1. Quality parameter assessment and data acquisition

The chickpea seed and its speckle image are shown in Fig. 3 (a) and (b) respectively. The QTP [30] were evaluated to reduce the subjectivity of the experimentation in the illumination parameters. Three quality protocols namely, saturation, contrast, and homogeneity were analysed and compared to obtain the best quality speckle images of the seeds. Saturation parameter indicates the under-exposure or saturation in speckle images by inferring pixel values of the images. These values should be in between 0 (under-exposure) to 255 (saturation) for accurate biospeckle analysis. The saturation map corresponding to the speckle image is shown in Fig. 4 (a). Contrast test is carried out to evaluate the integration period of the camera. This parameter is a direct measure of the speckle activity and a necessary step before the main experimentation. The mean contrast values are evaluated for the different time periods. These values should be in comparable range to maintain uniform contrast level throughout the experimentation. The bar graph for contrast parameter is shown in Fig.4 (b), which shows the significant correlation. Homogeneity test was performed to evaluate the level of homogeneity in the test object. Once all these parameters were evaluated and verified, the speckle images for different primed and un-primed seeds were captured.

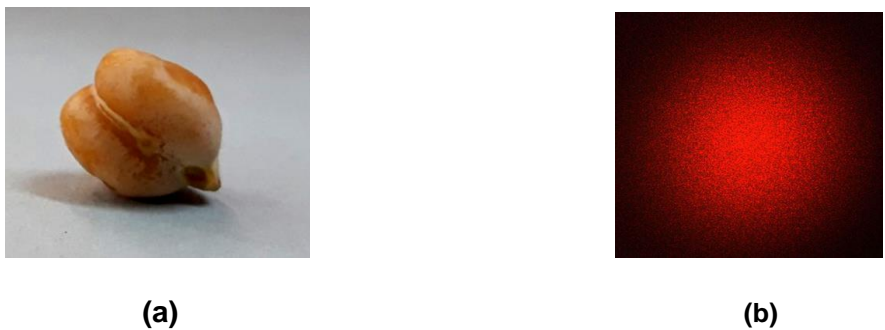

(a)                  (b)

**Fig. 3 (a) Chickpea seed, and (b) speckle image of the seed.**

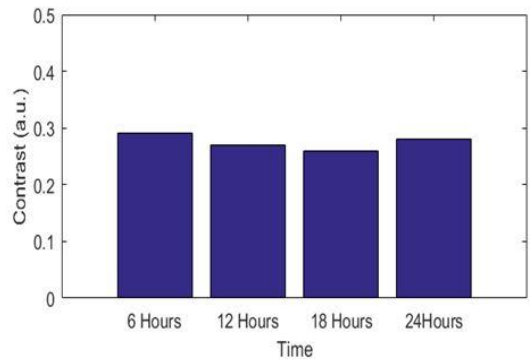

(a)                          (b)

**Fig. 4 (a) Saturation parameter, and (b) bar graph for contrast parameter.**

A collection of 100 consecutive images corresponding to each treated seed group ($T_0$-$T_{75}$) were collected (with the sampling rate of 0.05 seconds, the frame size of 1024×967 pixels) and stored in the computer memory. In order to ensure the accuracy of results, each experiment was repeated 5 times with 100 different seed samples. Next, pixel-by-pixel 'accumulation of differences' was performed using MSF technique within a window length of 5 frames. To access the statistical distribution of the quantitative data, a combination of non-normalized histogram mapping and Gaussian fitting was performed. Finally, in order to quantify the dynamicity of the primed seeds from the fitted histogram map, the frequency corresponding to the maximum histogram level was calculated. Position of the maximum in the fitted histogram data is the estimate of the underlying biospeckle activity.

4.2. Effect of hydro priming treatment

Priming method is based on the controlled seed hydration, close to the natural imbibition level; this allows the initiation of the seed germination process but inhibits the radicle perturbation [3]. This is an important process which affects the germination and seedling establishment of the crop. Hence, the specific metabolic changes in the seed after priming helps in water uptake and results in better germination characteristic. The result for the hydropriming treatment is shown in Fig. 5. The non-normalized histogram for the different treated seeds show the significant difference in the activity, as the biospeckle activity is directly related to the germination behaviour of the seed. The analysis of the speckle data shows substantial increase in the activity of hydro primed seed, indicating the improvement in germination traits. The highest biospeckle activity was achieved for the $T_4$ (hydropriming for 24 h). The germination rate for $T_1$ (hydropriming for 6 h) and $T_2$ (hydropriming for 12 h) was considerably lower as compare to $T_3$ and $T_4$. It was observed that, hydropriming for 24 h is most effective in increasing the germination characteristics for the chickpea seeds.

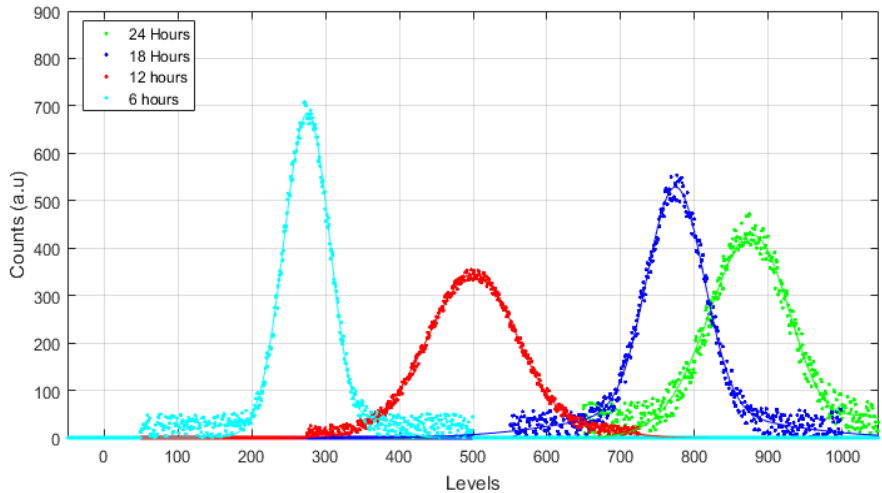

**Fig. 5 Histogram activity map of seeds subjected to hydropriming for different time periods.**

4.3. Effect of chemical priming treatment

Chemical priming is another method used for seed treatment. In this method, sodium chloride, potassium nitrate, and urea solutions were used with varying concentrations for different time periods, to treat the chickpea seeds. After priming with different chemical solutions, we observe that the seeds treated with potassium nitrate and urea shows very high biospeckle activity. By using solutions with low concentration, and for short duration, improved germination traits of the seed can be achieved. However, as the time duration or the concentration increases, the negative effect of priming starts, and the biospeckle activity decreases.

Fig. (6)-(8) show the results for the chemical priming using different solutions [$T_{51}$-$T_{75}$], the beneficial effect of the treatments are prominent for short period and low-concentration. These effects were attributed to the higher capacity for the osmotic adjustment in the treated seeds, metabolic healing process, and commencement of germination metabolites during the treatments. Priming with salt solution is effective for reducing the salinity and drought induced negative effect on the germination characteristic for different crops. The effectiveness of the short-term, low-concentration chemical priming on transcription at the initial phase of the seed

germination has been investigated in different plant species [31]. Kaya et al. [32] reported, the efficacy of the salt priming treatment under the saline and drought stress for sunflower seeds. The treatment enhances the seed germination, seedling emergence, and crop growth under the stressful environment. These beneficial effects of the treatment were observed due to enhanced metabolism of reverses as well as increase in endogenous abscisic acid (ABA) and gibberellin (GA) concentrations [30]. Furthermore, as the time or the concentration for treatment is increased, the osmotic potential decreases to a certain limit. This restricts the water uptake capability of the seed, thereby affecting the mobilization of nutrients necessary for the germination process. These metabolic changes negatively affects the germination characteristics of the seed. Increase in salt concentration also obstructs the water uptake capability of the seed from soil. This hampers seed imbibition process, resulting in increased time for the germination [6].

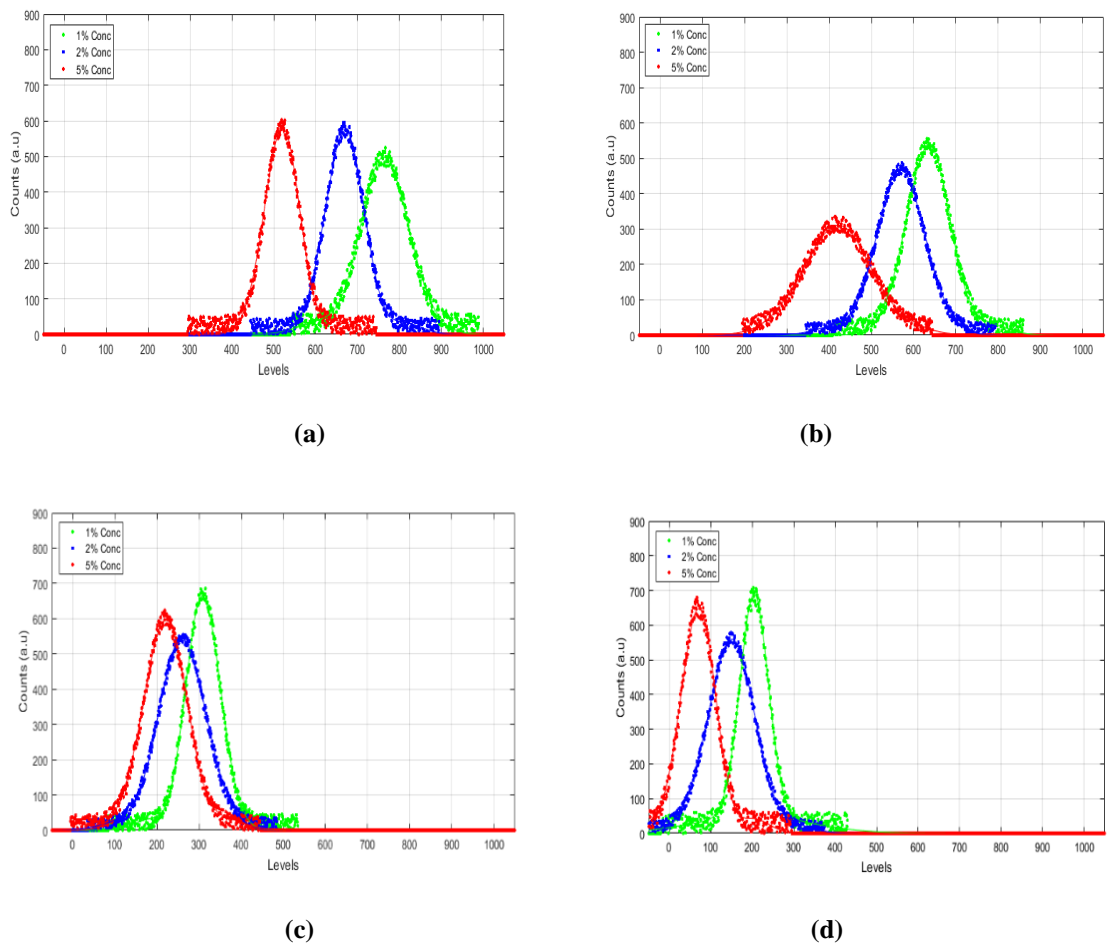

**Fig. 6. Histogram activity map of seeds subjected to chemical priming with varying NaCl concentration for different time periods (a) 6 Hours, (b) 12 Hours, (c) 18 Hours, and (d) 24 Hours.**

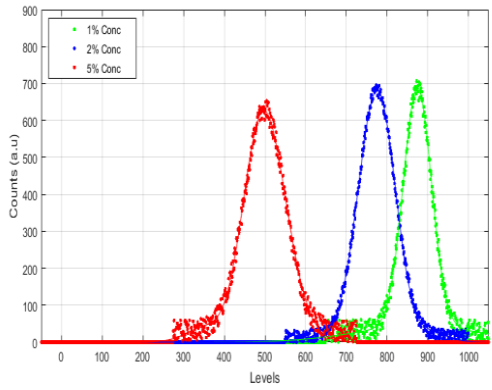
(a)

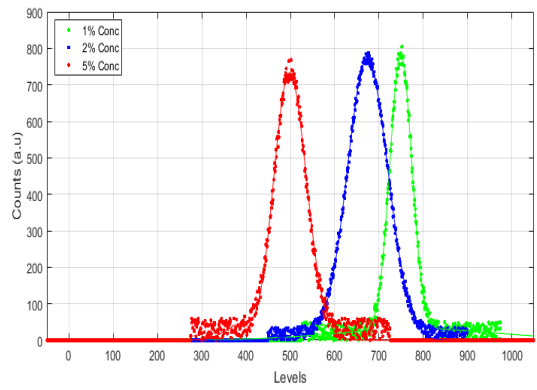
(b)

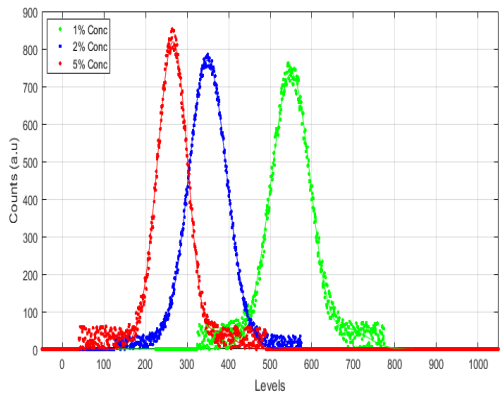
(c)

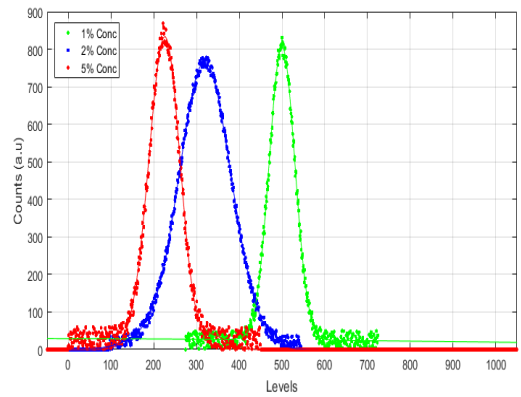
(d)

**Fig. 7. . Histogram activity map of seeds subjected to chemical priming with varying urea concentration for different time periods (a) 6 Hours, (b) 12 Hours, (c) 18 Hours, and (d) 24 Hours.**

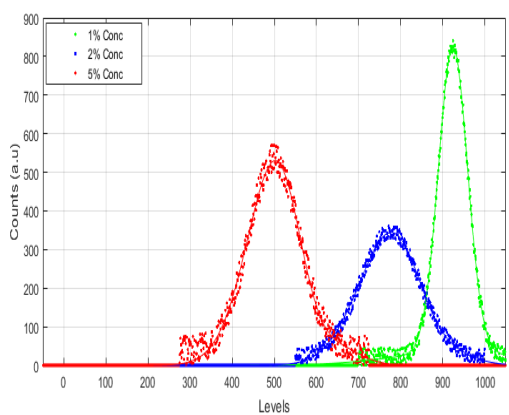
(a)

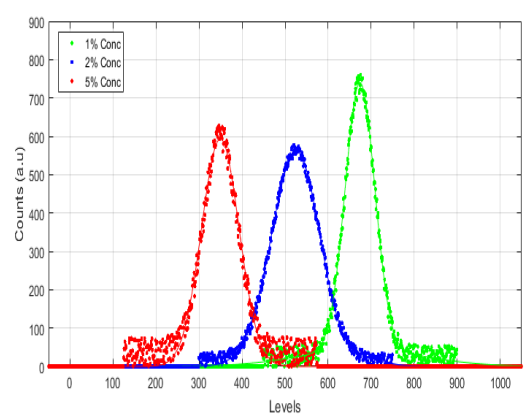
(b)

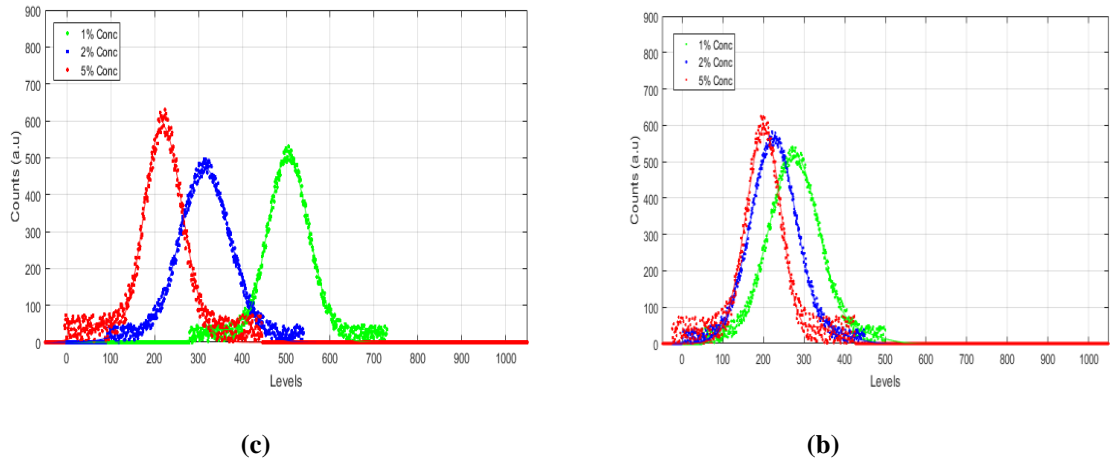

**Fig. 8. . Histogram activity map of seeds subjected to chemical priming with varying potassium nitrate concentration for different time periods (a) 6 Hours, (b) 12 Hours, (c) 18 Hours, and (d) 24 Hours.**

Key points for the comparison of different priming treatments are given below:

**(i) Chemical used:** Three chemical solutions, namely sodium chloride (NaCl), potassium nitrate ($KNO_3$), and urea ($CH_4N_2O$) were used for priming. Depending on efficiency of chemicals to enhance the germination characteristic of the seed, the solutions can be arranged in the following order:

$$KNO_3 > CH_4N_2O > NaCl$$

**(ii) Solution concentration:** Three different concentration of the chemicals i.e. 1%, 2%, and 5% were used for priming. The solution concentrations are arranged depending on the effectiveness to positively affect the germination behaviour of the seed:

$$1\%\ Conc > 2\%\ Conc > 5\%\ Conc$$

**(iii) Time duration of treatment:** The seeds were primed for four different time intervals, i.e. 6 hours, 12 hours, 18 hours, and 24 hours. According to the effectiveness on the germination parameters, the time period for effective priming can be arranged as:

$$6\ Hrs > 12\ Hrs > 18\ Hrs > 24\ Hrs$$

**4.4 Germination test**

In order to verify the results, we have conducted the germination test after completing the biospeckle analysis of the samples. The treated chickpea seeds were rinsed with the distilled water and air dried for 48 hours at 25°C (returned back to the initial moisture content); then instantly used for germination test. The test was performed in NSW-192 seed germinator (single chamber, 910mm x 555mm x 605 mm, humidity:

90 to 95% +/- 3%, temperature range: 5 to $60^0$ C). Seed germination test is used to access the capability of the seed to transform into a healthy plant under the desirable conditions necessary for the production process. The tests were performed under the laboratory conditions assuring specific light, moisture, and temperature parameters for the particular time period required for the germination process. The germination of a seed initiates with the imbibition process and ends up with the emergence of evident diffusion of the shoot above the soil surface. This process refers to the protrusion of a root and/or shoot from the seed coat.

The primed and un-primed seeds (5 replicates of 100 seeds/treatment) were subjected to the germination test and considered germinated when the radicles appear. The seeds were placed on germination papers, which was kept on watered petri dishes. These dishes were kept inside the germinator at $25\pm2^0$ C. To avoid the accumulation of salt, germination papers were replaced in every 48 hours. The germination percentage and mean germination time were calculated for all the treated seeds regularly up to day 7 after initiating the experiment [8]. The seeds were considered germinated when the 2 mm radicle emerged. The formula for germination percentage and mean germination time is given below:

Germination percentage $(GP) = \frac{Number\ of\ seeds\ germinated}{Total\ number\ of\ seeds\ used\ for\ the\ test} \times 100$

Mean germination time $(MGT) = \frac{\sum nD}{\sum n}$

where, $n$ is number of seeds germinated on day D, D is the days counted from the beginning of germination test, and the data is summed over the days of experimentation. Table 2 shows the correlation between germination parameters with biospeckle indices for all type of treated seeds.

**Table 2**

Comparison of germination traits with biospeckle activity.

| Priming treatments | Solution Concentration | Time for priming | | | | | | | | | | | |
|---|---|---|---|---|---|---|---|---|---|---|---|---|---|
| | | 6 Hours | | | 12 Hours | | | 18 Hours | | | 24 Hours | | |
| | | Biospeckle Index | Germination percentage (%) | Mean Germination time (days) | Biospeckle Index | Germination percentage (%) | Mean Germination time (days) | Biospeckle Index | Germination percentage (%) | Mean Germination time (days) | Biospeckle Index | Germination percentage (%) | Mean Germination time (days) |
| Non-primed | - | 4 | 84.25 | 7.52 | 4 | 84.25 | 7.52 | 0 | 84.25 | 7.52 | 4 | 84.25 | 7.52 |
| Hydropriming | - | 272.29 | 84.83 | 7.47 | 501.67 | 86.29 | 5.93 | 788.64 | 90.23 | 4.51 | 894.91 | 92.78 | 4.31 |
| Sodium chloride | 1% | 765.63 | 89.88 | 4.62 | 643.77 | 88.18 | 5.23 | 306.94 | 84.93 | 7.21 | 208.64 | 80.24 | 8.08 |
| | 2% | 678.27 | 88.42 | 5.04 | 567.85 | 87.93 | 5.61 | 261.74 | 82.23 | 7.88 | 142.38 | 79.82 | 8.17 |
| | 5% | 527.55 | 86.85 | 5.80 | 414.50 | 85.78 | 6.38 | 227.80 | 81.32 | 7.92 | 83.27 | 77.64 | 8.27 |
| Urea | 1% | 888.46 | 91.89 | 4.32 | 747.43 | 89.48 | 4.87 | 543.54 | 87.50 | 5.62 | 500.22 | 86.30 | 5.93 |
| | 2% | 789.24 | 90.63 | 4.49 | 684.42 | 88.63 | 4.92 | 447.18 | 85.97 | 6.02 | 323.95 | 85.02 | 7.02 |
| | 5% | 502.04 | 86.37 | 5.97 | 498.46 | 86.27 | 5.92 | 264.56 | 82.25 | 7.83 | 231.12 | 81.37 | 7.89 |
| Potassium nitrate | 1% | 913.58 | 93.09 | 4.29 | 678.24 | 88.43 | 5.03 | 504.07 | 86.40 | 5.97 | 267.19 | 83.78 | 7.82 |
| | 2% | 789.56 | 90.78 | 4.43 | 527.19 | 86.87 | 5.81 | 312.35 | 84.98 | 7.19 | 207.07 | 80.23 | 8.08 |
| | 5% | 497.19 | 86.21 | 5.94 | 342.87 | 85.24 | 6.87 | 218.34 | 81.02 | 7.98 | 298.61 | 84.92 | 7.23 |

The values are rounded off to the nearest integer for the given numerical data.

**5. Conclusion**

In this paper, dynamic speckle phenomenon has been investigated as a potential tool for analyzing the effect of priming on chick pea seed. Highlights of the proposed setup include its fast speed of operation, extreme simplicity, low cost, the requirement of lesser number of components, simple algorithms and its ability to be commercialized. Following conclusions can be drawn from this study:

I. We have developed a direct method to access the effect of different seed priming treatments using biospeckle analysis without prior germination. As the effect of priming is analyzed directly, the speed of operation of the proposed sensor in considerably higher as compared to all other existing techniques.

II. Biospeckle indexing of the seed priming behavior includes a combination of MSF, non-normalized histogram and Gaussian fitting. The algorithm processes the biospeckle activity map in visual, statistical and numerical domains. As a result, it provides accurate and robust assessment of priming behavior as compared to other existing techniques of analysis.

III. We investigated the effect of different priming treatments on germination traits using different chemical solutions (sodium chloride, potassium nitrate and urea), concentration (1%, 2%, and 5%) and time duration (6 hours, 12 hours, 18 hours, and 24 hours). It was observed that, hydropriming for 24 hours and chemical priming with $KNO_3$ for lower concentration (1%) and short time duration (6 hours) are most effective for improving the overall germination characteristics of chick pea seeds.

IV. The proposed strategy is very simple, non-destructive, easy-to-align, low cost and full field.

Further investigations are underway to develop a commercial product using laser biospeckle to determine the effect of priming on different seeds. Furthermore, the effect of different parameters e.g. temperature, humidity, aging. on priming will also be analyzed using the same technique for the improvement of crop yield.

**Acknowledgement**

This publication is an outcome of the R&D work undertaken project under the Visvesvaraya PhD Scheme of Ministry of Electronics & Information Technology, Government of India, being implemented by Digital India Corporation, and Department of Science & Technology project grant (EMR/2016/003115/EEC).

9. S. Paparella, S. S. Araujo, G. Rossi, M. Wijaysinghe, D. Carbonera, and A. Balestrazzi, "Seed priming: state of art and new perspectives," Plant Cell Rep 34, 1281–1293 (2015).
10. F. Waller, B. Achatz, H. Baltruschat, J. Fodor, K. Becker, M. Fischer, T. Heier, R. Huckelhoven, C. Neumann, and G. Von-Wettstein, "The endophytic fungus Piriformis indica reprograms barley to salt-stress tolerance, disease resistance, and higher yield," Proc Nat Acad Sci102, 13386–13391 (2005).
11. Keher Singh, "Laser speckles: fundamental properties and some applications," Light and its many wonders, Viva Books Private Limited, 449-472 (2016).
12. A. Chatterjee, V. Bhatia, and S. Prakash, "Anti-spoof touchless 3D fingerprint recognition system using single shot fringe projection and biospeckle analysis," Opt. Las. Engg. 95, 1-7 (2017).
13. A. Chatterjee, P. Singh, V. Bhatia, and S. Prakash, "A low cost optical sensor for secured antispoof touchless palm print biometry," IEEE Sensors Letters 2, 1-4 (2018).
14. M. Pajueloa, G. Baldwina, H. Rabal, N. Cap, R. Arizagab, and M.Trivib, "Bio-speckle assessment of bruising in fruits," Optics and Lasers in Engineering 40, 13-24 (2003).
15. F. M. Vincitorio, N. Budini, C. Mulone, M. Spector, C. Freyre, A. J. L. Diaz, and A. Ramil, "Detection of fungi colony growth on bones by dynamic speckle," Proc. of SPIE, 8754-8754D21-6 (2013).
16. R. A. Braga, L. Dupuy, M. Pasqual, and R. R. Cardosos, "Live biospeckle laser imaging of root tissues," Euro. Biophys. Journ. 38, 679–686 (2009).
17. G. D. S. Guedes, K. T. M. Guedes, D. R. Dias, R. F. Schwan, and R. A. Braga, "Assessment of biological activity of Kefir grains by biospeckle laser technique," Afr. Journ. Microbe. Res. 8, 2639-2642 (2014).
18. I. C. Amaral, R. A. Braga, E. M. Ramos, A. L. S. Ramos, and E. A. R. Roxael, "Application of biospeckle laser technique for determining biological phenomena related to beef aging," Journ. Food Engg. 119, 135-139 (2013).
19. A. Chatterjee, R. Disawal, and S. Prakash, "Biospeckle Assessment of Bread Spoilage by Fungus Contamination Using Alternative Fujii Technique", Proc. Of International conference on Opto-Electronics and Applied optics (OPTRONICS-2016) Kolkata 18-20, (2016).
20